\documentclass[amsmath,amssymb,prl,reprint]{revtex4-1}
\usepackage{graphicx}
\usepackage{dcolumn}
\usepackage{bm}
\usepackage{color}
\definecolor{dickred}{rgb}{0,0,0}
\setcitestyle{super}

\begin{document}

\title{Proximity effect and interface transparency in Al/InAs-nanowire/Al diffusive junctions}
\author{A.V.~Bubis}
\affiliation{Institute of Solid State Physics, Russian Academy of
Sciences, 142432 Chernogolovka, Russian Federation}
\affiliation{Moscow Institute of Physics and Technology, Dolgoprudny, 141700 Russian Federation}
\author{A.O.~Denisov}
\affiliation{Institute of Solid State Physics, Russian Academy of
Sciences, 142432 Chernogolovka, Russian Federation}
\affiliation{Moscow Institute of Physics and Technology, Dolgoprudny, 141700 Russian Federation}
\author{S.U.~Piatrusha}
\affiliation{Institute of Solid State Physics, Russian Academy of
Sciences, 142432 Chernogolovka, Russian Federation}
\affiliation{Moscow Institute of Physics and Technology, Dolgoprudny, 141700 Russian Federation}
\author{I.E.~Batov}
\affiliation{Institute of Solid State Physics, Russian Academy of
Sciences, 142432 Chernogolovka, Russian Federation}
\affiliation{Moscow Institute of Physics and Technology, Dolgoprudny, 141700 Russian Federation}
\author{J.~Becker}
\affiliation{Walter Schottky Institut, Physik Department, and Center for Nanotechnology and Nanomaterials, Technische Universit\"{a}t M\"{u}nchen, Am Coulombwall 4, Garching 85748, Germany}

\author{J.~Treu}
\affiliation{Walter Schottky Institut, Physik Department, and Center for Nanotechnology and Nanomaterials, Technische Universit\"{a}t M\"{u}nchen, Am Coulombwall 4, Garching 85748, Germany}

\author{D.~Ruhstorfer}
\affiliation{Walter Schottky Institut, Physik Department, and Center for Nanotechnology and Nanomaterials, Technische Universit\"{a}t M\"{u}nchen, Am Coulombwall 4, Garching 85748, Germany}

\author{G.~Koblm\"{u}ller}
\affiliation{Walter Schottky Institut, Physik Department, and Center for Nanotechnology and Nanomaterials, Technische Universit\"{a}t M\"{u}nchen, Am Coulombwall 4, Garching 85748, Germany}
\author{V.S.~Khrapai} 
\affiliation{Institute of Solid State Physics, Russian Academy of
Sciences, 142432 Chernogolovka, Russian Federation}
\affiliation{Moscow Institute of Physics and Technology, Dolgoprudny, 141700 Russian Federation}

\begin{abstract} 
We investigate the proximity effect in InAs nanowire (NW) junctions with superconducting contacts made of Al. The carrier density in InAs is tuned by means of the back gate voltage $V_g$. At high positive $V_g$ the devices feature transport signatures characteristic of diffusive  junctions with highly transparent interfaces -- sizable excess current, re-entrant resistance effect and proximity gap values  ($\Delta_N$) close to the Al gap ($\Delta_0$). At decreasing $V_g$, we observe a reduction of the proximity gap down to $\Delta_N\approx\Delta_0/2$ at NW conductances $\sim2e^2/h$, which is interpreted in terms of carrier density dependent reduction of the Al/InAs interface transparency. We demonstrate that the experimental behavior of $\Delta_N$ is closely reproduced by a model with shallow potential barrier at the Al/InAs interface. 
\end{abstract}

\maketitle

A possibility to interface a superconductor (S) with a metallic (N) state of InAs-based mesoscopic semiconductors is known since 
at least two decades~\cite{denHartog1996,Lachenmann1997}. High critical and excess currents, sometimes comparable to a theoretical upper limit,
are reliably observed in InAs nanowire (NW) Josephson junctions~\cite{Nishio2011,Abay2012,Paajaste2015}. Observations of signatures of a Cooper pair splitting~\cite{Hofstetter2009,Das2012} further conform with high device quality. A recent revival of interest in proximity effect in S-NW devices emerged in the course of a search for Majorana zero-energy states~\cite{Alicea2012}, with the research targeted at single conduction channel NWs~\cite{Das2012_ZBA,Mourik2012,Deng2016,Albrecht2016}. 

Thanks to a well-known Fermi level pinning effect, which results in a charge accumulation at the surface of InAs~\cite{Olsson1996,Speckbacher2016}, the NWs grown from this material typically exhibit low contact resistance to normal metals. For the superconducting hybrid devices, however, even a weak interface reflectivity can sufficiently degrade the performance by suppressing the Andreev reflection in favor of the normal quasiparticle scattering~\cite{Octavio1983,Flensberg1988}. Although the transparency of the S/NW interfaces can be controlled by various chemical treatments prior to the deposition of the superconductor~\cite{Nishio2011,Abay2012,Paajaste2015,Gl2017} or, alternatively, via the {\it in-situ} MBE growth of the superconductor~\cite{Chang2015}, residual imperfections are still present~\cite{Abay2014,Kjaergaard2017}. One might expect that the underlying physics can be clarified by tracking the carrier density dependence of the proximity effect, since the effective interface transparency should critically depend on the shape of the potential barrier.

In this work we study the evolution of the superconducting proximity effect in an InAs NW with evaporated Al contacts as a function of electron density in InAs, which is controlled by means of a global back gate. At high carrier densities, the transport data is compatible with highly transparent Al/InAs interfaces and the induced proximity gap ($\Delta_N$) is close to the superconducting gap value in Al ($\Delta_0$). The observed reduction of $\Delta_N$ by a factor of two at decreasing carrier density indicates gate-tunable interface transparency, which can be explained assuming a shallow potential barrier formed at the Al/InAs interface.
\begin{figure}[t]
\begin{center}
\vspace{10mm}
  \includegraphics[width=0.8\linewidth]{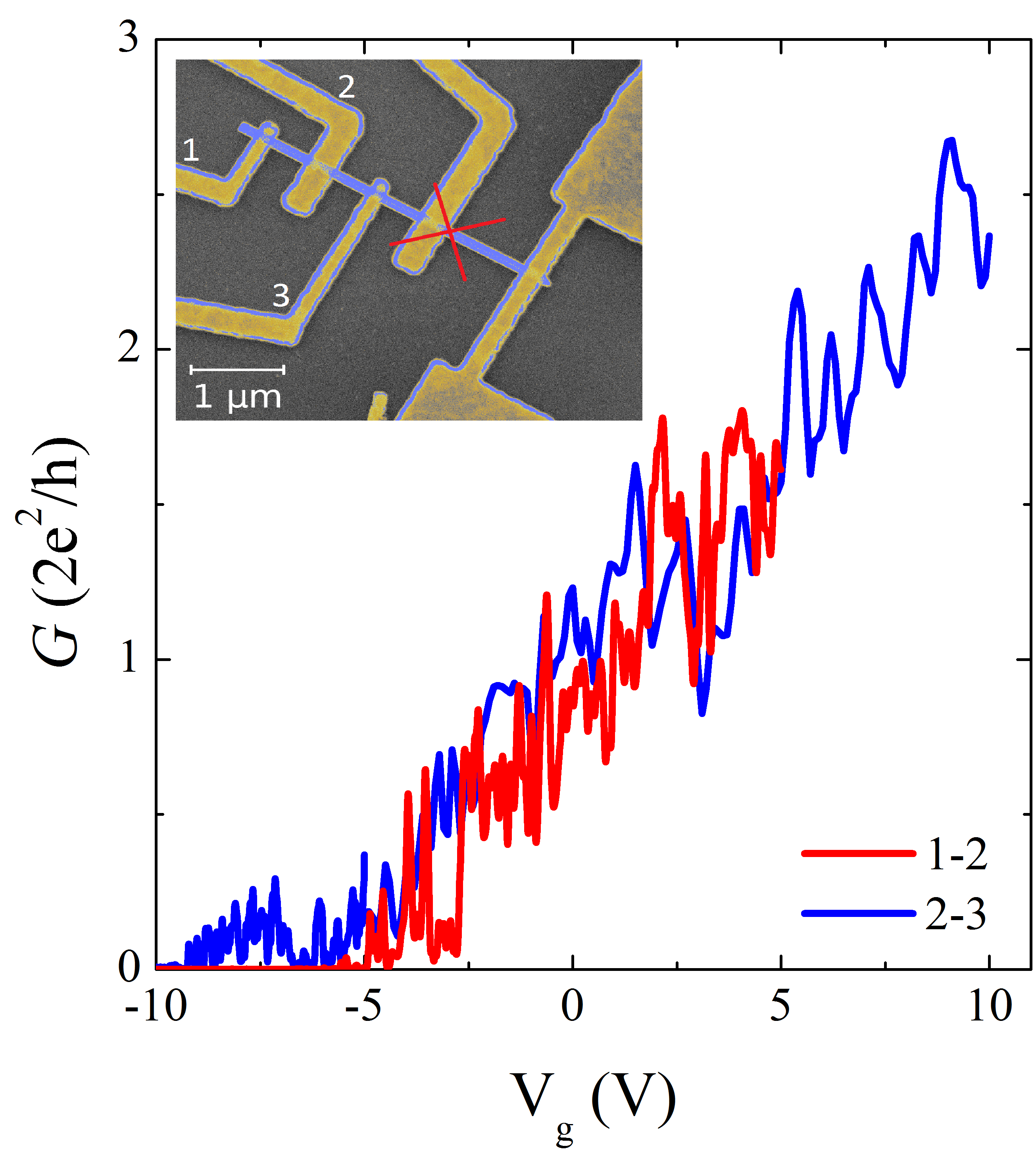}
\end{center}
  \caption{Conductance of the NW. Main figure: gate voltage dependencies of the two-terminal linear response NW conductance, taken at 0.5\,K and $B=50\,$mT for the two contact configurations, see legend. Inset: false color SEM image of the device identical to the one used throughout the paper. The contacts 1, 2 and 3 were used for NW measurements. The Al stripe in the right bottom corner was used for the reference measurements in Al.} 
	\label{fig1}
\end{figure}

\begin{figure*}[t]
\begin{center}
\vspace{0mm}
  \includegraphics[bb = 0.1cm 1cm 22cm 17.5cm, clip=true, height=1.\columnwidth]{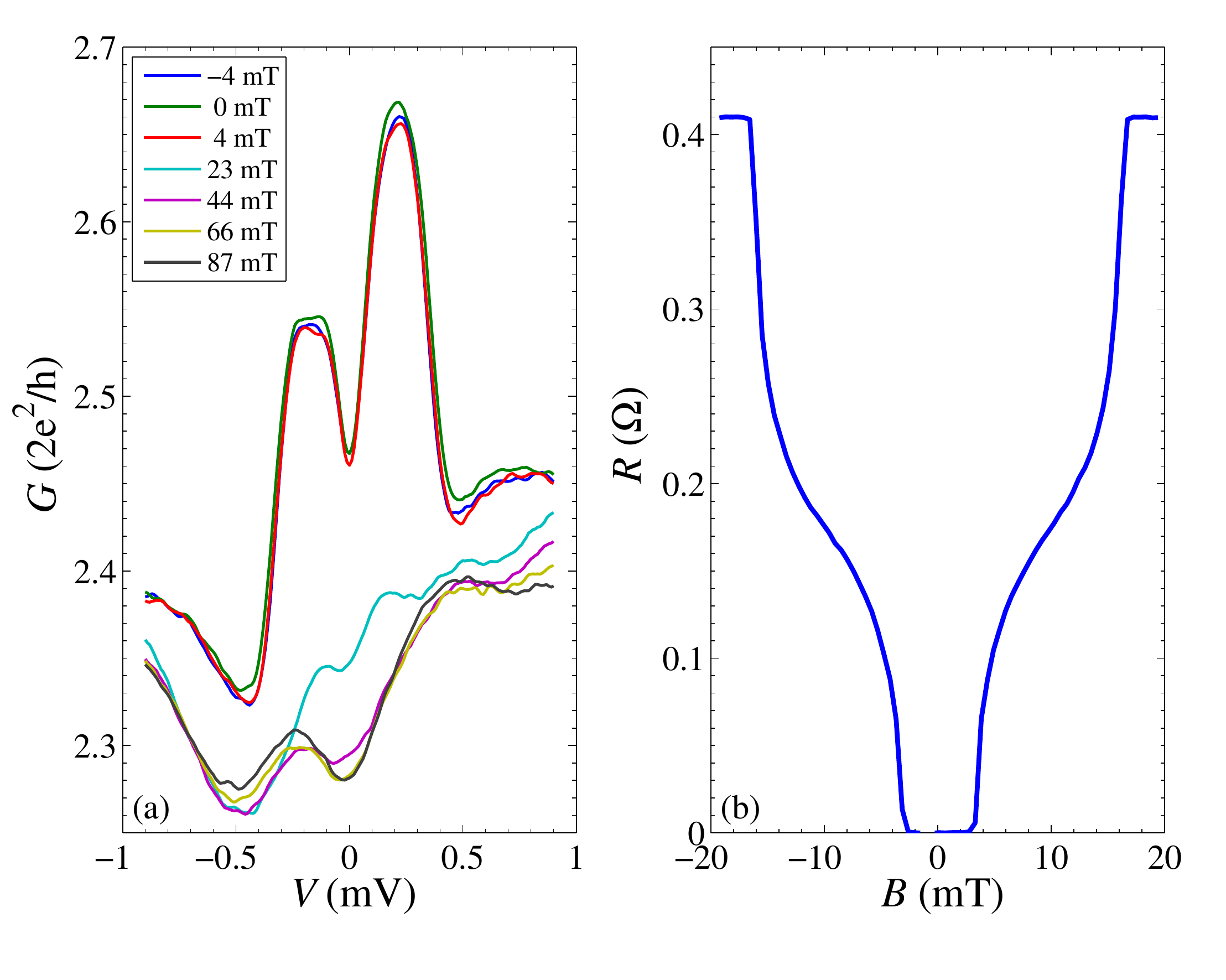}
	\includegraphics[bb = 0cm -0.1cm 6cm 11.65cm, clip=true, height=1.\columnwidth]{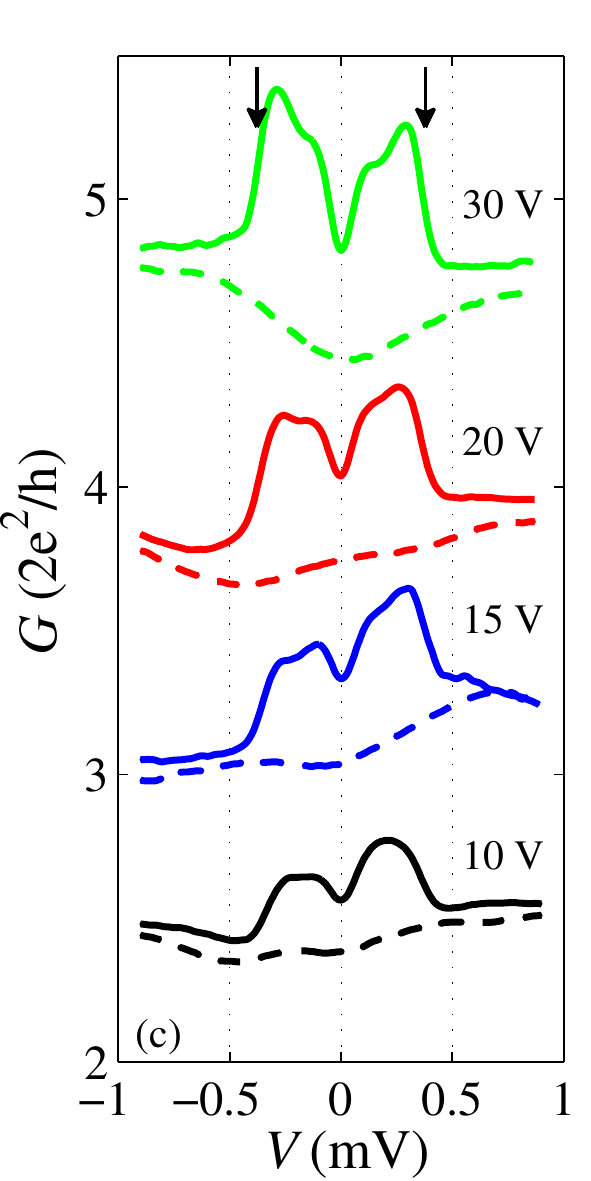}
		   \end{center}
  \caption{Influence of voltage bias and magnetic field on the superconducting proximity effect. (a): Differential conductance of the NW as a function of $V$ for a set of $B$-field values, see legend. Proximity effect results in an increase of $G$, that can be suppressed by a small perpendicular magnetic field. $G$ features, which result from the superconductivity (small $B$) and mesoscopic oscillations (higher $B$) are discussed in the text. (b): $B$-dependence of the four terminal resistance of the Al reference contact, see the bottom right contact in the inset of Fig.~\ref{fig1}. Narrow Al stripe becomes superconducting below $\approx3\,$mT and the normal state is fully restored above $\approx18\,$mT. (c): bias voltages dependence of $G$ for several $V_g$ values (legend) in $B=0$ (solid lines) and $B=50\,$mT (dashed lines). Vertical arrows mark the positions $|V|=2\Delta_0/e$. The NW data in (a) and (c) were taken for the contact configuration 1-2.}
	\label{fig2}
\end{figure*}

We investigate unintentionally doped catalyst free InAs NWs grown by MBE on a  Si(111) substrate. The NWs were removed from the substrate in an ultrasonic bath and drop casted on a piece of commercial $\rm n^+-$doped Si wafer covered by 300\,nm of $\rm SiO_2$. Following a standard e-beam lithography process and prior to an electron beam evaporation of Al (150\,nm at $1\times10^{-7}\,$mbar) the sample was  {\it in-situ} exposed to Ar ion gun (4.5 sccm Ar flow, 40 V discharge and 400 V accelerating voltage during 4 min) in order to remove the native oxide layer from the contact regions. The evaporation was performed in a Plassys MEB 550S system. Transport measurements were performed in a $\rm ^3He$ insert at a base temperature of 0.5\,K using a home-made current-voltage converter and a lock-in amplifier, with an ac bias modulation in the range of 10-15\,$\mu$V rms at frequencies 17-33\,Hz. Altogether we investigated two devices, one of which had a very weak capacitive coupling to the back gate and didn't permit a wide range tuning of the conductance. In the following we concentrate on the data obtained from the second device.

A false color SEM image of an Al-InAs NW device identical to the measured one is shown in the inset of Fig.~\ref{fig1}. The InAs NW and five ohmic contacts (lighter colors) are placed on the $\rm SiO_2/Si$ substrate (darker color). The right-most contact was shaped as a 300\,nm wide and 2\,$\rm \mu m$ long Al reference channel and had two pairs of terminals on either side. This contact was used for reference transport measurements of Al presented below. The critical temperature of the superconducting phase transition measured in the reference channel is $T_c=1.20\pm0.03\,$K, corresponding to the superconducting energy gap of $183\pm5\,\mu$eV in Al contacts. Below we use a slightly higher gap value of $\Delta_0=190\,\mu$eV, which allows a better fit of the proximity gap data in InAs. The two NW sections marked by a cross were not conducting, presumably owing to the electrostatic discharge, leaving three contacts 1, 2 and 3 available for transport measurements. The device depicted in Fig.~\ref{fig1} precludes unambiguous evaluation of the Al/InAs interface resistance. Nevertheless, we have checked that, down to the gate voltage $V_g=0$, the three- and two-terminal resistances are almost identical, up to the wiring contribution of $\sim100\,\rm \Omega$. It follows that the equilibration length of the chemical potential difference across the Al/InAs interface is well below the width of the contact 2 (360 nm). Below we discuss only the results of the two-terminal conductance measurements.

In the main Fig.~\ref{fig1} we plot the $V_g$ dependence of the linear response conductance, $G$, for the two NW sections. 
For both contact configurations, see legend, the conductance exhibits an overall increase at increasing $V_g$ with reproducible mesoscopic fluctuations. Sizable conductance $G\sim2e^2/h$ around $V_g=0$ is a consequence of the Fermi level pinning at the surface of InAs, which results in surface charge accumulation in our nominally undoped NWs~\cite{Olsson1996,Speckbacher2016}. Roughly linear $V_g$ dependence of $G$ is consistent with previous measurements on these and other similar NWs~\cite{Abay2014} . The onset of NW conduction allows to roughly estimate~\cite{Wunnicke2006} the NW depletion point (the threshold voltage) at $-10\,V< V_{th}< -5\,V$. Within a simplified 3D charge carrier model this corresponds to the charge carrier density of $2\div4\times10^{17}{\rm cm^{-3}}$ in the un-gated NW. The estimates of the mobility, $\sim1000{\rm cm^{2}/Vs}$, and the mean-free path, $l\sim20{\rm nm}$ at $V_g=0$ are consistent with preliminary measurements on the NWs from the same batch with Ni/Au contacts.

In Fig.~\ref{fig2}a we demonstrate the impact of the magnetic field, $B$, directed perpendicular to the substrate, on the NW differential conductance, $G=dI/dV$. Here, we plot the bias, $V$, dependence of $G$ for contacts 1 and 2 at $V_g=10$\,V. For $|B|>25$\,mT, the $G$ is almost independent of the magnetic field, and the observed weak bias dependence comes from the evolution of the mesoscopic fluctuations. Consistent with this interpretation, we observed that in this $B$ range the $G(V)$ traces  exhibit both the local maxima and minima depending on the $V_g$ choice, see dashed lines in Fig.~\ref{fig2}c for more data. In small magnetic fields and within the bias window $|V|\lesssim200\,\mu$V the $G$ increases by $\sim10$\% , nearly independent of $B$ for $|B|\lesssim4$\,mT. This behavior correlates with the four-terminal resistance, $R$, measurements of the Al reference channel. As shown in Fig.~\ref{fig2}b, the Al film becomes superconducting ($R=0$) in small magnetic fields. At increasing $|B|$ above $\gtrsim3$\,mT the resistance gradually increases until reaching the normal state value above 18\,mT. The step-like $B$ dependence of $R$ originates presumably from a smaller critical field of Al in the narrow channel as compared to much wider pads (see the inset of Fig.~\ref{fig1}). We conclude that the increase of $G$ in small $B$ in Fig.~\ref{fig2}a is a consequence of the superconducting proximity effect induced by Al contacts in the InAs NW.

In Fig.~\ref{fig2}c we compare the $G(V)$ traces for the Al contacts in a superconducting state ($B=0$, solid lines) and in a normal state ($B=50\,$mT, dashed lines) at large positive gate voltages $V_g\geq10\,$V. Here, nearly independent of $V_g$, a sharp increase of $G$ is observed at $B=0$ around the bias voltage roughly corresponding to twice the superconducting energy gap of the Al film $e|V|\approx2\Delta_0=380\,\mu$eV, see vertical arrows. This is similar to previously reported diffusive Al/InAs-NW based SNS junctions~\cite{Nishio2011,Abay2012}. Around zero bias, the conductance exhibits a sizable local dip, reminiscent of a well-known re-entrant resistance behavior~\cite{Artemenko1979,Volkov1996,Nazarov1996,Golubov1997} observed in normal metal~\cite{Kozhevnikov2000} and semiconductor~\cite{Lachenmann1997} based NS junctions. In diffusive SNS junctions, the width of the dip $\sim50\,\mu$eV reflects the value of the proximity minigap, $\varepsilon_g$, and the Thouless energy, $\varepsilon_T\equiv\hbar D/L^2$ in the NW ($\varepsilon_g\approx3.2\varepsilon_T$ for highly transparent intefaces~\cite{Cuevas2006}), consistent with the independent estimate of $\varepsilon_T\sim15\,\mu$eV. The tiny dips around $e|V|\approx\Delta_0$ resolved on the upper three traces in Fig.~\ref{fig2}c are the second MAR features. Altogether, the data of Fig.~\ref{fig2} indicate a high quality of Al/InAs interfaces in our devices at large positive $V_g$, similar to the state of the art Al/InAs-NW devices with evaporated Al~\cite{Nishio2011,Abay2014}. 

\begin{figure}[t]
\begin{center}
\vspace{10mm}
  \includegraphics[width=0.8\linewidth]{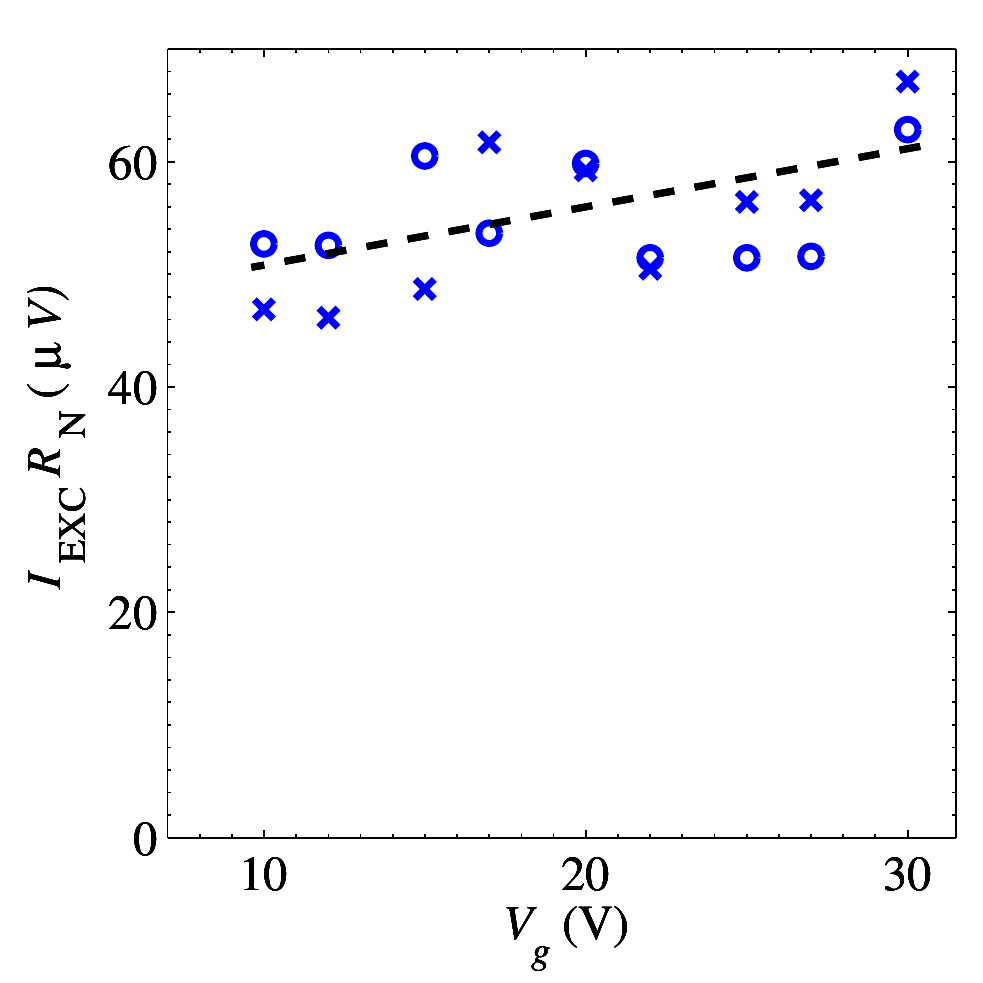}
\end{center}
  \caption{Excess current. $V_g$ dependence of the product $I_{exc}R_N$ calculated from the conductance data for the NW contacts 1-2.
	The data for the positive bias voltages is represented by circles, the data for the negative $V$ by crosses. The dashed line is a guide to the eye. The reduction of $I_{exc}R_N$ compared to the theoretical upper limit is a joint effect of the reduced interface transparency and the incoherent transport regime, see text.}
	\label{fig3}
\end{figure}

\begin{figure*}[t]
\begin{center}
\vspace{10mm}
  \includegraphics[width=0.9\linewidth]{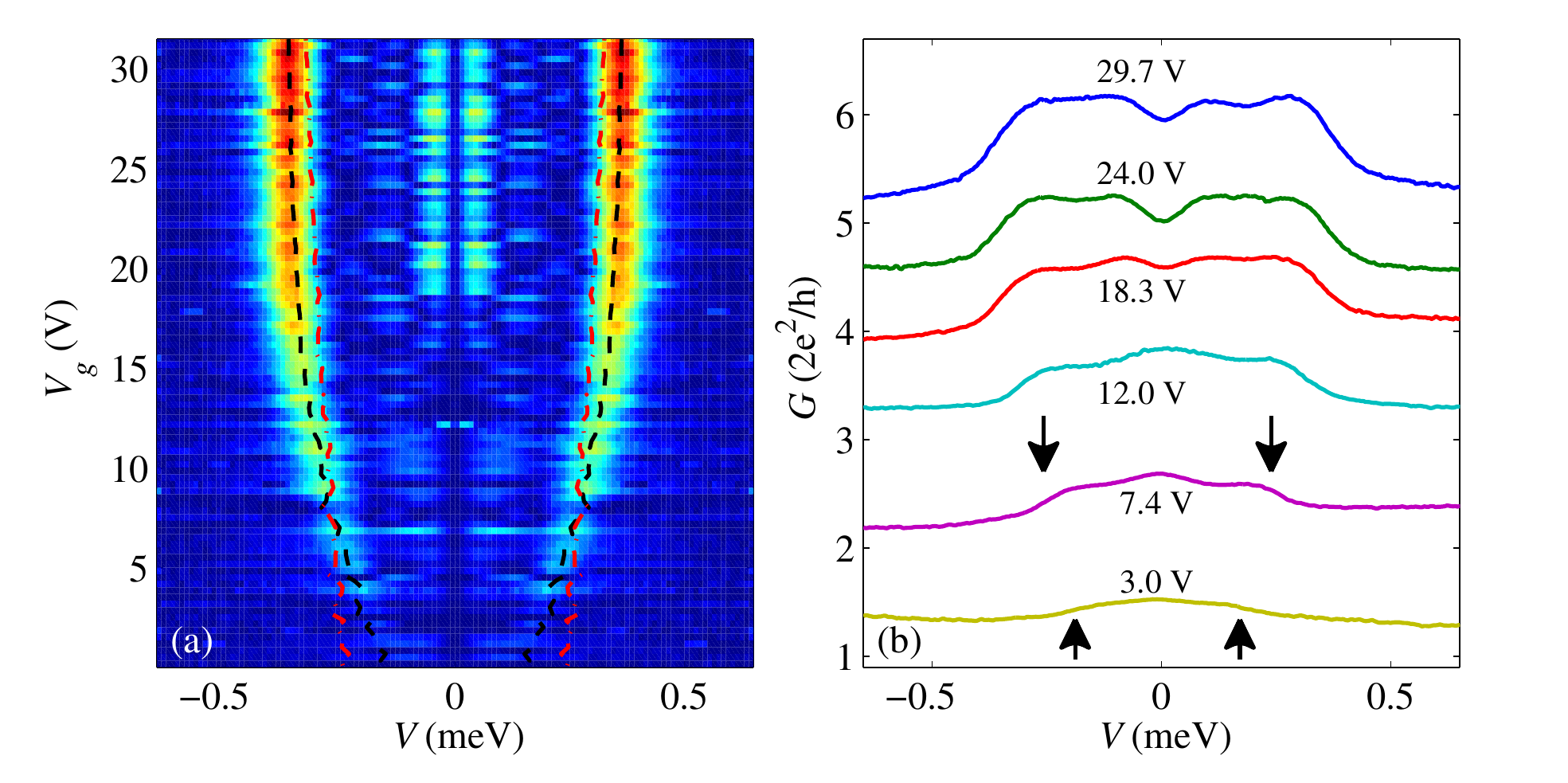}
\end{center}
  \caption{Gate voltage dependence of the proximity gap induced in InAs. (a): color scale plot of the absolute value of the derivative $|dG/dV|$ as a function of $V_g$ and $V$. The two brightest lines correspond to the sharp change of $G$ at the bias positions of the doubled proximity gap $V=\pm2\Delta_N/e$. The dashed line and the dash-dotted line are the fits obtained, respectively, for the model of shallow rectangular potential barrier at the Al/InAs interface and for the BTK $\delta$-shaped potential model. (b): representative traces $G$ vs $V$ for several gate voltages. The positions $V=\pm2\Delta_N/e$ for the two lowest traces are marked by vertical arrows. All the data is obtained at $B=0$ for the contacts 2-3, the data in (a) were symmetrized with respect to the bias reversal.}
	\label{fig4}
\end{figure*}

We gain further insight about the transparency of the Al/InAs interfaces in our SNS junctions from the analysis of the excess current, defined as 
\begin{equation*}
	 I_{exc}=\int_0^{|V_{high}|}{\left[G(B=0)-G(B)\right]}d|V|,
\end{equation*}
where the differential conductances $G(B=0),\,G(B)$ are measured at zero and high enough magnetic fields, respectively, and the upper bias limit is $|V_{high}|>2\Delta_0/e$. In Fig.~\ref{fig3} we plot the gate voltage dependence of the product $I_{exc}R_N$ for $V_{high}=\pm800\,\mu$V, where $R_N$ is the linear response normal state resistance of the NW. As already expected from the data of Fig.~\ref{fig2}, we observe a positive excess current. Within the experimental uncertainty, $I_{exc}R_N$ is independent of the bias polarity and the observed data scatter is a consequence of mesoscopic fluctuations. Overall, in Fig.~\ref{fig3} we observe a reduction of $I_{exc}$ at decreasing $V_g$, which occurs faster than the increase of $R_N$. The highest measured values of the excess current reach $eI_{exc}\approx R_N^{-1}\Delta_0/3$, which is still a factor of 5 lower than the ideal value of $eI_{exc}\approx(\pi^2/4-1)R_N^{-1}\Delta_0$ in a diffusive coherent SNS junction~\cite{Artemenko1979,Bardas1997} or the value $1.05R_N^{-1}\Delta_0$ in a short double barrier SINIS junction~\cite{Brinkman2000}. The reduction of $I_{exc}$ can originate from the interface scattering and/or the loss of coherence in InAs. In our devices, the ratio of the superconducting coherence length in InAs, $\xi=(\hbar D/\Delta_0)^{1/2}$ and the junction length $L$ is $\xi/L\sim1/2$, which cannot alone explain the effect~\cite{Cuevas2006}. We conclude that the role of the Al/InAs interfaces is important for the observed reduction of $I_{exc}$, similar to Ref.~\cite{Abay2014}, where comparable values of $I_{exc}$ in long junctions were observed. 

A close inspection of the bias dependent $G$ over the extended gate voltage interval further clarifies the origin of the 
interface scattering. In Fig.~\ref{fig4}a we show a color-scale plot of the derivative $|dG/dV|$ as a function of both $V$  and $V_g$. This data was obtained for the junction between the contacts 2 and 3. The most pronounced feature is the sharp change of $G$, which manifests itself as two wide bright lines for positive and negative bias voltages in Fig.~\ref{fig4}a. The bias positions of the lines, $|V|\approx2\Delta_0/e$, are almost independent of the gate voltage for $V_g\gtrsim15\,$V. At lower $V_g$, however, the lines start to converge rapidly and become fainter, which corresponds to the weakening of the $G$ features. This behavior is detailed in Fig.~\ref{fig4}b, where the selected traces $G(V)$ are plotted for various $V_g$. For instance, the traces at $V_g=7.4\,$V and $V_g=3\,$V exhibit conductance steps at, roughly, $|V|\approx1.4\Delta_0/e$ and $|V|\approx\Delta_0/e$, respectively, see the vertical arrows. Conductance features in NS, as well as SNS, junctions are a complex function of a two-gap density of states in such systems~\cite{Volkov1993,Aminov1996,Zaitsev1998}, one being the superconductor energy gap, $\Delta_0$, and the other, $\Delta_N$, the proximity induced gap in the N region. In the following, we relate  the positions of the discussed $G$ features with the values of the proximity gap induced in InAs underneath the Al contacts, $|V|=2\Delta_N/e$. The data of Fig.~\ref{fig4}, therefore, demonstrates that $\Delta_N$ is a strong function of the charge carrier density in the NW.  

The diminished proximity gap is a consequence of the reduced transparency of the Al/InAs interfaces, as determined by an implicit relation~\cite{Aminov1996}:
\begin{equation}
		\Delta_N=\frac{\Delta_0}{1+0.56\gamma_B\sqrt{1-(\Delta_N/\Delta_0)^2}}.
		\label{eq_aminov}
\end{equation}

Here, $\gamma_B\equiv\sigma_{N}/(G_I\xi_N)$ is the interface parameter, which is determined by the InAs conductivity $\sigma_N$, the coherence length $\xi_N=(\hbar D/\Delta_0)^{1/2}$, where $D$ is the diffusion coefficient in InAs, and the interface conductance per unit area, $G_I$. The latter is expressed by a generalized Sharvin formula, $G_I=(e^2/h)\left\langle T\right\rangle k_F^2/2\pi$, where $k_F$ is the Fermi wave vector in InAs and $\left\langle T\right\rangle$ is the angle averaged transmission probability of the interface. In the following, we compare the evolution of the conductance features in Fig.~\ref{fig4} with the behavior of $\Delta_N$ for two models of the interface scattering potential.

The inevitable source of scattering at the interface of two different materials lies in a mismatch of their Fermi velocities~\cite{Schaepers_book}. The case of Al and InAs is special in that the small Fermi momentum in InAs is compensated by the small electronic effective mass $\approx0.023m_e$, as compared to Al. This gives rise in almost ideally matched Fermi velocities at carrier densities on the order of $N_e=10^{18}\,{\rm cm^{-3}}$ in InAs. We estimate that in our devices this source of scattering results in $\left\langle T\right\rangle\geq0.7$ (and $\gamma_B\ll1$) for $V_g>0$ and thus can be safely neglected. 

The conventional approach to the interface scattering is based on a BTK formalism~\cite{BTK1982}, which postulates a $\delta$-shaped tunnel barrier at the interface. The strength of the barrier is characterized by a dimensionless parameter $Z=Z_0/\cos{\theta}$, where 
$Z_0\propto1/v_F$, $v_F$ is the Fermi velocity in InAs and $\theta$ is the angle of incidence with respect to the normal to the interface. The angle-averaged transmission probability, $T\equiv 1/(1+Z^2)$, is given by $\left\langle T\right\rangle=1-Z_0^2\ln{(1+Z_0^{-2})}$. Using eq.~(\ref{eq_aminov}) we calculate the gate voltage dependence of the proximity gap and plot it in Fig.~\ref{fig4}a. The best fit to the conductance features at $V=\pm\Delta_N/e$ was obtained assuming $Z_0\approx1.4$ at $N_e=1\times10^{18}\,{\rm cm^{-3}}$ (dashed-dotted lines). The fit captures well the overall reduction of $\Delta_N$ predicting a nearly monotonic $V_g$ dependence, but fails to closely reproduce the experimental behavior in Fig.~\ref{fig4}. This observation, however, can be tackled assuming a shallow rectangular potential barrier of height $U$ and thickness $t$ formed at the Al/InAs interface. The $V_g$ dependence of $\Delta_N$ for such a barrier is shown by the dashed lines in Fig.~\ref{fig4}a. This fit is obtained for $U=90\,$meV and $t=13\,$nm and closely reproduces the experimental behavior of conductance features. The saturation of $\Delta_N\approx\Delta_0$ occurs in this case well above the $V_g\approx4\,$V, which corresponds to a match of the Fermi energy and the classical turning point of the barrier, $E_F=U$. 

\begin{figure}[t]
\begin{center}
\vspace{10mm}
  \includegraphics[width=0.8\linewidth]{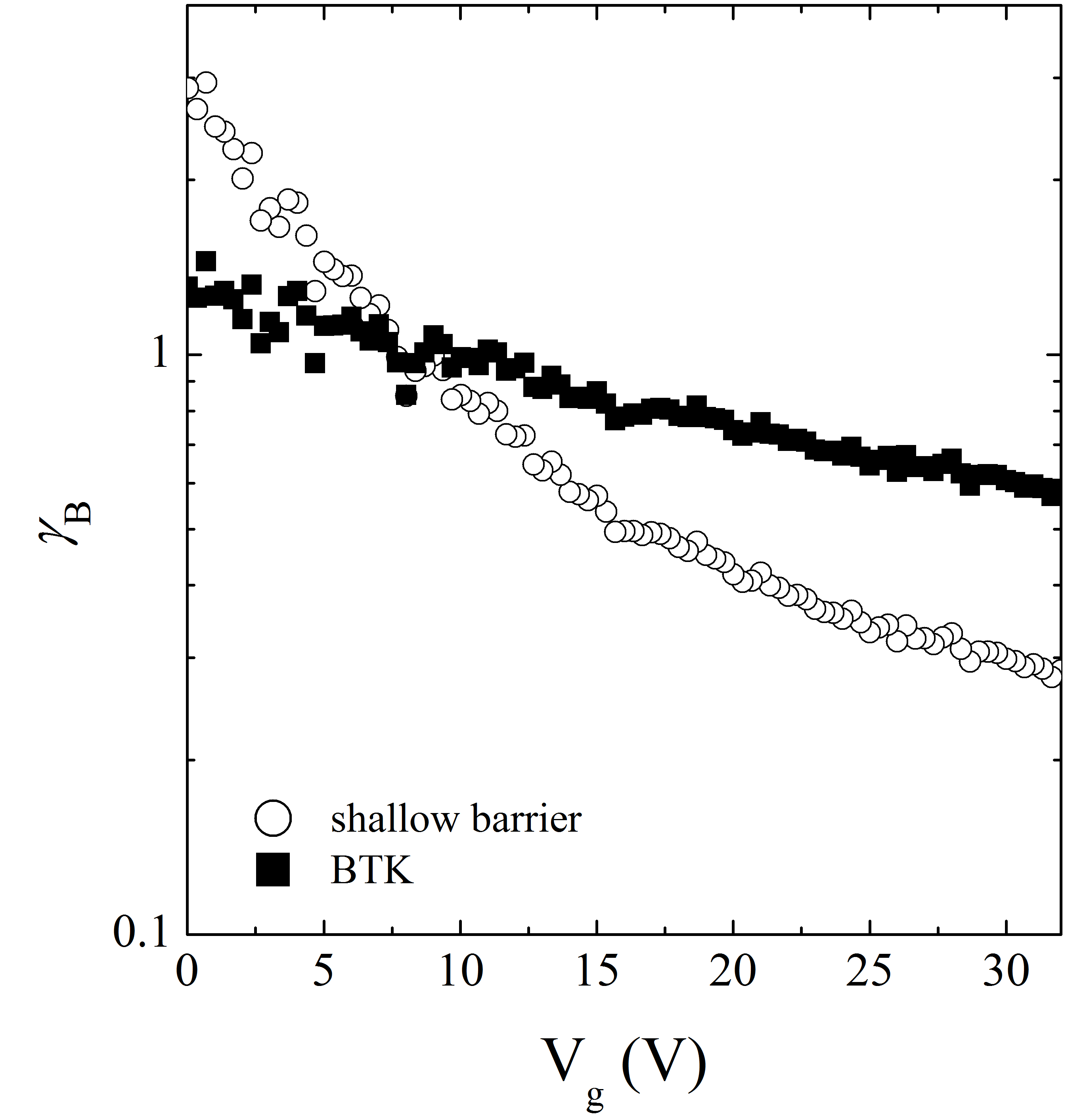}
\end{center}
  \caption{Interface scattering parameter. Gate voltage dependencies of the parameter $\gamma_B$, calculated for the two models of the Al/InAs interface potential (legend) and used to fit the data of Fig.~\ref{fig4}a. }
	\label{fig5}
\end{figure}

The difference between the two models of the interface barrier is highlighted in Fig.~\ref{fig5}, where we plot the $V_g$ dependencies of the interface parameter $\gamma_B$ used to fit the data of Fig.~\ref{fig4}a. The much stronger dependence for the case of rectangular barrier (open circles) is explained by the exponential suppression~\cite{Landau3} of $\left\langle T\right\rangle$ for $E_F\ll U$, in contrast to the power law $E_F$ dependence in the BTK case. Note, that for the rectangular barrier and high positive $V_g$ the model predicts an almost ideal interface transparency with $\gamma_B\approx0.3$, which might be difficult to correlate with the observation of $I_{exc}R_N<\Delta_0/e$ in Fig.~\ref{fig3}. We expect that a slightly higher value of $\gamma_B\sim1$ would be more realistic in our device in the high carrier density regime, similar to a recent report in epitaxial Al/InAs heterostructures~\cite{Kjaergaard2017}. 

The physical possibility of some kind of a shallow barrier at the Al/InAs interface seems natural. On the one hand, a strong insulating temperature dependence in NWs with Al contacts at negative gate voltages, which is observed even in the highest quality structures~\cite{Chang2015}, is qualitatively compatible with a thermal activation over such barrier. The formation of schottky barriers of the height on the order of 100\,meV were reported roughly independent of the NW diameter~\cite{Razavieh2014} and for various metals~\cite{Feng2016}. On the other hand, the barrier thickness of $t=13\,$nm,  used to describe the data of Fig.~\ref{fig4}, is the same order of magnitude as the Bohr radius in InAs ($\sim35$\,nm), which sets a natural length scale for the band bending. 

In summary, we investigated the proximity effect and its gate voltage dependence in Al/InAs NW/Al diffusive junctions. At high positive $V_g$ several differential conductance features indicate highly transparent Al/InAs interfaces. However, at decreasing $V_g$ the proximity effect weakens and the value of the proximity gap, induced underneath the Al contacts, reduces by almost a factor of two compared to the Al superconducting gap. This observation is interpreted in terms of the carrier density dependent transparency of the Al/InAs interfaces, which is best described by assuming that a shallow potential barrier is present at the interface. Our results clarify possible origin of the interface scattering in Al/InAs NWs devices, suggesting that a control of the carrier density in the contact region might be helpful for future applications.

We gratefully acknowledge discussions with A.A.\,Golubov, Ya.V.\,Fominov, S.\,Ludwig and E.S.\,Tikhonov. This work was supported by the RSF-DFG Project No. 16-42-01050 and, in part, by the RSF Project No. 15-12-30030 and the Ministry of Education and Science of the Russian Federation Grant No. 14Y.26.31.0007.  We thank D.V.\,Negrov and E.V.\,Korostylev for a technical assistance and for an access to the equipment of MIPT Center of Collective Usage.
\bibliography{Bubis}

\end{document}